%% file: main.tex
\documentclass[10pt,twocolumn,letterpaper]{article}

\usepackage{iccv}              

\input{preamble}

\definecolor{iccvblue}{rgb}{0.21,0.49,0.74}
\usepackage[pagebackref,breaklinks,colorlinks,allcolors=iccvblue]{hyperref}
\usepackage{hyperref} 

\def\paperID{8992} 
 \def\confName{ICCV}
\def\confYear{2025}

\title{RnGCam: High-speed video from rolling \& global shutter measurements}

\author{
Kevin Tandi$^{1}$\thanks{Joint first authors}\quad
Xiang Dai$^{1*}$ \quad
Chinmay Talegaonkar$^{1}$ \quad
Gal Mishne$^{1,2}$ \quad
Nick Antipa$^{1}$ \quad \\
$^{1}$Department of Electrical and Computer Engineering, University of California, San Diego \\
$^{2}$ Halıcıoğlu Data Science Institute, University of California, San Diego \\
{\tt\small \{ktandi, xidai, ctalegaonkar, gmishne, nantipa\}@ucsd.edu}
}

\begin{document}
\maketitle
\input{sec/0_abstract}    
\input{sec/1_intro}

\input{sec/2_related_works}
\input{sec/3_method}

\input{sec/4_experimental_setup}
\input{sec/5_implementation}
\input{sec/6_results}

\input{sec/7_limitations}

\input{sec/8_future_work}

{
    \small
    \bibliographystyle{ieeenat_fullname}
    \bibliography{main}
}

\input{sec/X_suppl}
\end{document}

%% file: preamble.tex
\newcommand{\psf}{h}

\newcommand{\crop}{\text{C}}
\newcommand{\pad}{\text{P}}
\newcommand{\meas}{b}
\newcommand{\obj}{v}
\newcommand{\objVec}{\mathbf{\obj}}
\newcommand{\objVecTheta}{\mathbf{\obj}_{\theta}}
\newcommand{\measVec}{\mathbf{\meas}}
\newcommand{\Ars}{\mathbf{A}_R}
\newcommand{\Ags}{\mathbf{A}_G}
\newcommand{\UVec}{\mathbf{U}}
\newcommand{\Afull}{\mathbf{A}}

\usepackage{float}

%% file: sec/0_abstract.tex
\begin{abstract} 

Compressive video capture encodes a short high-speed video into a single measurement using a low-speed sensor, then computationally reconstructs the original video. Prior implementations rely on expensive hardware and are restricted to imaging sparse scenes with empty backgrounds. We propose RnGCam, a system that fuses measurements from low-speed consumer-grade rolling-shutter (RS) and global-shutter (GS) sensors into video at kHz frame rates. The RS sensor is combined with a pseudorandom optic, called a diffuser, which spatially multiplexes scene information.
The GS sensor is coupled with a conventional lens. The RS-diffuser provides low spatial detail and high temporal detail, complementing the GS-lens system's high spatial detail and low temporal detail.
We propose a reconstruction method using implicit neural representations (INR) to fuse the measurements into a high-speed video. Our INR method separately models the static and dynamic scene components, while explicitly regularizing dynamics. In simulation, we show that our approach significantly outperforms previous RS compressive video methods, as well as state-of-the-art frame interpolators. We validate our approach in a dual-camera hardware setup, which generates 230 frames of video at 4,800 frames per second for dense scenes, using hardware that costs $10\times$ less than previous compressive video systems. 

\end{abstract}

%% file: sec/1_intro.tex
\section{Introduction}
\label{sec:intro}

High-speed video imaging is instrumental in visualizing and analyzing fast-moving systems across disciplines, such as neuroscience~\cite{fournely2018high, lu2017video} and microscopy~\cite{zhou2023parallelized, ishijima1995high}. Conventional image sensors have limited analog-to-digital bandwidth, which limits the spatio-temporal sampling rate a given sensor can acquire. This forces a trade-off between temporal and spatial resolution. 
Conventional high-speed cameras use expensive, bulky sensors and read architectures to reduce the trade-off by directly increasing bandwidth.

In contrast, \textit{compressive video} breaks this trade-off by encoding multiple frames of high-speed video into a single digital exposure captured with a 2D image sensor. 
The video frames are then computationally reconstructed. 
The resulting inverse problem is ill-posed and requires strong video priors to uniquely recover the video~\cite{antipa2019video, Weinberg2020100000FC}. 

\begin{figure}
    \centering
    \includegraphics{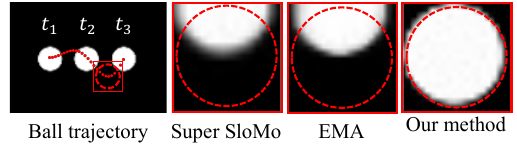}
    \caption{\textbf{Video interpolators struggle with complex motions.} Even with a very simple background, video interpolators produce inaccurate trajectories of the ball, with the 3 input frames at $t_1,t_2,$ and $t_3$. Our method recovers these trajectories with high fidelity.}
    \label{fig:toy_example_sineball}

\end{figure}

While many hardware solutions for compressive video have been proposed, we focus here on exploiting the rolling shutter (RS) available in nearly all low-cost CMOS image sensors. Previous work shows that high-speed compressive video can be recorded by coupling RS sensors with optical multiplexing elements such as diffusers \cite{antipa2019video, Weinberg2020100000FC, sheinin2018rolling}.  However, these methods rely on sparse video priors.
As a result, they struggle to recover dense scenes with bright, detailed backgrounds. This limits their utility to relatively simple scenes with empty backgrounds, which are consistent with the sparsity priors.  
Data-driven video interpolation methods~\cite{jiang2018super,lu2022video,kalluri2023flavr} are commonly used to upsample videos captured by conventional low-fps cameras. However, since these methods are typically trained on internet videos, they often generalize poorly to out-of-distribution scenarios, particularly when interpolating chaotic motions over large temporal gaps. We demonstrate this with a simple toy experiment in~\cref{fig:toy_example_sineball}.
We aim to accelerate conventional sensors by over $100\times$, a regime in which learned upsampling performs poorly for chaotic, out-of-distribution scenes. 
\paragraph{Our Contributions:}In this paper, we address these limitations and demonstrate that RS sensors can capture high-speed compressive video of scenes comprising nontrivial backgrounds, by leveraging a few GS frames captured during the exposure of the RS sensor. We build a hardware prototype, \textit{RnGCam}, to capture optically aligned GS and RS measurements. 
We propose an implicit neural representation (INR)-based space-time fusion model (STFM), to recover high-speed videos from the combination of diffuser-coded RS and GS measurements. Using GS measurements and our proposed regularization, we recover high-speed videos with dense backgrounds with much higher fidelity than previous methods~\cite{antipa2019video, 10378452}. We demonstrate our improvements over previous work, and modern data-driven video interpolators~\cite{jiang2018super, zhang2023extracting} in both simulation and on real-world data captured from \textit{RnGCam}. 

\begin{figure*}[h]
    \centering
     \includegraphics[width=1\linewidth]{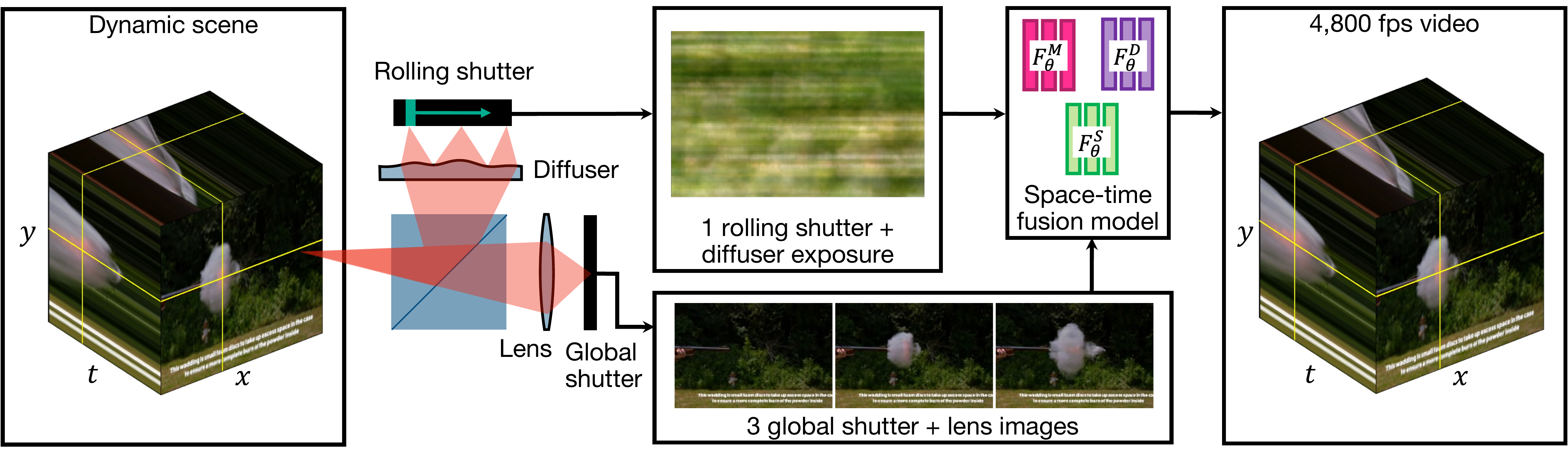}
    \caption{\textbf{Pipeline for fusing multiple global shutter measurements and an RS diffuser coded long exposure measurement.} Both sensors are triggered at the same time, and in between the start and end of the RS's coded long exposure, two more images are captured as key frames using the global shutter. The RS and diffuser encodes high speed dynamics into a single measurement, and the GS measurements act as key frames for the reconstruction. The sum of a time-varying and static neural scene representation is used to fuse together both measurements into a high-speed reconstruction with a dense background. }
    \label{fig:Pipeline}
\end{figure*}

The rest of the paper is organized as follows. We start with related work in \cref{sec:related_works}, and outline the camera model preliminaries in \cref{sec:camera_model}. We present our space-time fusion model (STFM) reconstruction algorithm in \cref{sec:neural_ST_model}. We demonstrate results on simulated and real data, and explain the RnGCam hardware setup in \cref{sec:results}.

%% file: sec/2_related_works.tex
\section{Related work}
\label{sec:related_works}

Methods for high-speed video recovery from regular sensors can be broadly classified into two categories. \textit{Hardware modulation} methods, which use spatial and temporal multiplexing, or capture data with additional sensors such as event cameras, to complement standard sensors. 
\textit{Computational Methods}, which rely on algorithmic techniques such as data-driven video interpolation methods, or INRs to solve the inverse problem of video upsampling from compressive measurements. 





\noindent \textbf{Spatial-Multiplexing Systems:  }
Enhancing the temporal resolution of an imaging sensor can be achieved through spatial multiplexing using optical elements such as diffraction gratings or diffusers placed in front of the sensor. These elements map a single scene point to multiple sensor pixels, allowing each pixel to carry scene information across different time frames. 
Spatial multiplexing allows video recording with a limited number of pixels, reducing the bandwidth requirements. 
Different pixel arrangements have been used to subsample the sensor plane, including a single pixel \cite{duarte2008single}, a line sensor ~\cite{sheinin2020diffraction}, region of interest (ROI) ~\cite{sheinin2021deconvolving}, or a conventional RS sensor ~\cite{antipa2019video,Weinberg2020100000FC,sheinin2018rolling}. 
Because RS sensors read row-by-row, each row can encode a frame of video, increasing the frame rate by a factor proportional to the number of rows.
In all cases, reconstructing a full-frame image from limited pixel measurements is an ill-conditioned problem. 
Many of these works which rely on sparsity priors, inspired by compressed sensing, perform poorly for non-sparse scenes.  
Some work has shown improvements using INRs to enforce stronger priors, improving results significantly~\cite{monakhova2021untrained, Cao2024}.
In our work, we show that our hardware, comprising two complementary cameras, in conjunction with our INR-based video reconstruction method, produces significantly better video quality for dense scenes than previous spatial multiplexing systems. 

\noindent \textbf{Temporal-Multiplexing Systems:  }
Coding various exposures within a single frame has been widely utilized in capturing dynamic scenes to deblur the motion artifacts~\cite{holloway2012flutter}. Later, this concept was extended to reconstructing high-speed video from a single measurement ~\cite{veeraraghavan2010coded, hitomi2011video, martel2020neural, 5585094,sankaranarayanan2013compressive,iliadis2020deepbinarymask, harmany2011spatio}. 
A straightforward method is to modulate exposure time pixel-wise using a specially designed sensor that offers single-pixel exposure control~\cite{martel2020neural, iliadis2020deepbinarymask,portz2013random, willett2011compressed}.
Another approach requires dynamic optical components such as a streak camera~\cite{gao2014single}, piezoelectric stages~\cite{koller2015high, liu2018rank,llull2013coded}, spatial light modulator (SLM)~\cite{hitomi2011video,sankaranarayanan2013compressive, reddy2011p2c2} or digital micromirror device (DMD)~\cite{deng2019sinusoidal,wang2017compressive} to generate different designed patterns at a higher time rate to encode the temporal information. 
These specialized optical components and sensors are expensive and have a limited frame rate. 
Additionally, many of the methods struggled in non-sparse scenes~\cite{sheinin2021deconvolving, sheinin2020diffraction}.
Our method uses a homemade diffuser made of optical epoxy on cover slides, while still allowing us to recover frame rates of up to $4800$ with a single calibrated PSF image.

\noindent \textbf{Multi-Sensor systems:}
Another common approach is combining multiple sensor types to compensate for the limitations of a single sensor. Event cameras, valued for dynamic sensitivity, capture information missed by regular RGB sensors~\cite{tulyakov2021time, tulyakov2022time, zhang2024event}. However, they are expensive, saturate with camera motion, and Timelens \cite{tulyakov2021time} requires training with a large video-event dataset, and reports at most 15 frame skip compared to our 130. Similarly, camera arrays have achieved high-speed video capture~\cite{wilburn2005high}. Other works have fused RS and GS for video recovery without optical multiplexing, recovering video with relatively low frame rates ~\cite{fan2021inverting}. Integrating consumer-grade GS and RS sensors to collect complementary data is an innovative approach to decoding high-speed information~\cite{Sheinin:2022:Vibration}, though it has not been directly applied to video recovery. Building on this idea, our method inserts a diffuser in front of the RS to encode dynamic information instead of a speckle pattern while using a standard RGB sensor to preserve high-frequency details.

\noindent \textbf{Data-driven video frame interpolation methods:  } Our work closely relates to video frame interpolation (VFI), a well-studied computer vision problem ~\cite{parihar2022comprehensive}. Popular VFI methods use convolution- or transformer-based models with implicit architectural priors and are trained on large datasets. Recent works predict frames as discrete time steps given a start and end frame \cite{Li_2023_CVPR, kalluri2023flavr, niklaus2017video, lu2022video}. In contrast, ours enables continuous frame interpolation along the time axis. We hence compare our work with the popular method SuperSlomo~\cite{jiang2018super}, and more recent work EMA-VFI~\cite{zhang2023extracting}, which allow continuous time interpolation between frames. Our approach outperforms these methods (see \cref{sec:results}), without being trained on large datasets.



\noindent \textbf{Implicit Neural Representations (INRs) for inverse problems:  }
Coordinate-based neural networks have been gaining popularity for a variety of visual computing tasks, for a full survey, refer to \cite{10.1111:cgf.14505}. Our work closely relates to approaches~\cite{Cao2024, 9924573,9606601} that use INRs in computational imaging inverse problems. In our paper, we adopt SIRENs \cite{DBLP:journals/corr/abs-2006-09661} as our signal representation. A variety of other neural representations have been developed with different architectures to increase the representation ability of these coordinate networks \cite{Saragadam_2023_CVPR,lindell2021bacon, Xie_2023_CVPR}.

%% file: sec/3_method.tex
\section{Fusing GS and coded RS images to handle complex scenes}
\label{sec:camera_model}

In an optical system that utilizes spatio-temporal multiplexing optics like ours, there is a finite amount of information that can be encoded into a single measurement. This is particularly apparent in complex scenes with stronger backgrounds, where the background signal is mixed and aggregated with the signal of interest and makes the recovery of the dynamic portion of the signal much more challenging. 

As illustrated in Figure \ref{fig:Pipeline}, our system consists of two sensors aimed at the same scene by utilizing a beam splitter to divide the light equally between the coded RS arm and the arm with the GS sensor at imaging conditions.

In this section, we first present the acquisition scheme for our data capture. We then illustrate the image formation model for the coded RS image and GS images.
Our goal in this paper, is to exploit the RS behavior to gain access to the data that GS would miss in the long gaps between frames, while using the high quality GS images, imaged with a conventional lens, to improve the spatial detail of the recovered high-speed video.  
\subsection{General camera model}
Here we describe the model for a general sensor combined with an optical system with a known shift-invariant PSF, $h(\xi,\eta)$. Assuming no occlusions, the time-varying intensity arriving at the sensor from a dynamic scene $\obj(\xi,\eta,\tau)$ is given by 2D linear convolution
\begin{equation}\label{eq:linear_conv}
    \widetilde{\obj}(\xi,\eta,\tau)=\obj\left(\xi,\eta,\tau\right)\stackrel{(\xi, \eta)}{*}h(\xi,\eta),
\end{equation}
where $\stackrel{(\xi,\eta)}{*}$ denotes 2D convolution over the continuous spatial dimensions $(\xi,\eta)$, for time $\tau$.  We discretize the problem onto grid $(x,y,t)$ and approximate \eqref{eq:linear_conv} as
\begin{equation}\label{eq:discrete_2D_conv}
    \widetilde{\obj}(x,y,t) = \crop \left[ \pad \left( \psf(x,y) \right) \stackrel{(x,y)}{\circledast} \pad \left( \obj(x,y,t)\right ) \right]
\end{equation}
where $\circledast$ is circular convolution, $\pad$ operator is 2D zero padding, and $\crop$ is 2D cropping such that $\crop(\pad(v)) = v$. 
Note that our optical system includes a field stop which limits the support of the scene to an area strictly smaller than the sensor, which is why $\obj$ can be zero-padded in \eqref{eq:discrete_2D_conv}. Finally, the digital measurement recorded by a sensor is 
\begin{equation}\label{eq:general_exposure}
    \meas(x,y) = \sum_t S(x,y,t)\widetilde{\obj}(x,y,t){,}
\end{equation}
where $S(x,y,t)$ is an indicator that encodes the sensor's exposure timing at pixel $(x,y)$, taking on value 1 when a pixel is actively recording photons, and 0 otherwise. Substituting \eqref{eq:discrete_2D_conv} into \eqref{eq:general_exposure} yields a general camera model for a system with a known PSF and sensor exposure pattern. In subsequent sections, we describe $S$ and $h$ for each of our two cameras. 

\subsection{Rolling and Global shutter timings}
\begin{figure}[h!]
    \centering
    \includegraphics[width=1\linewidth]{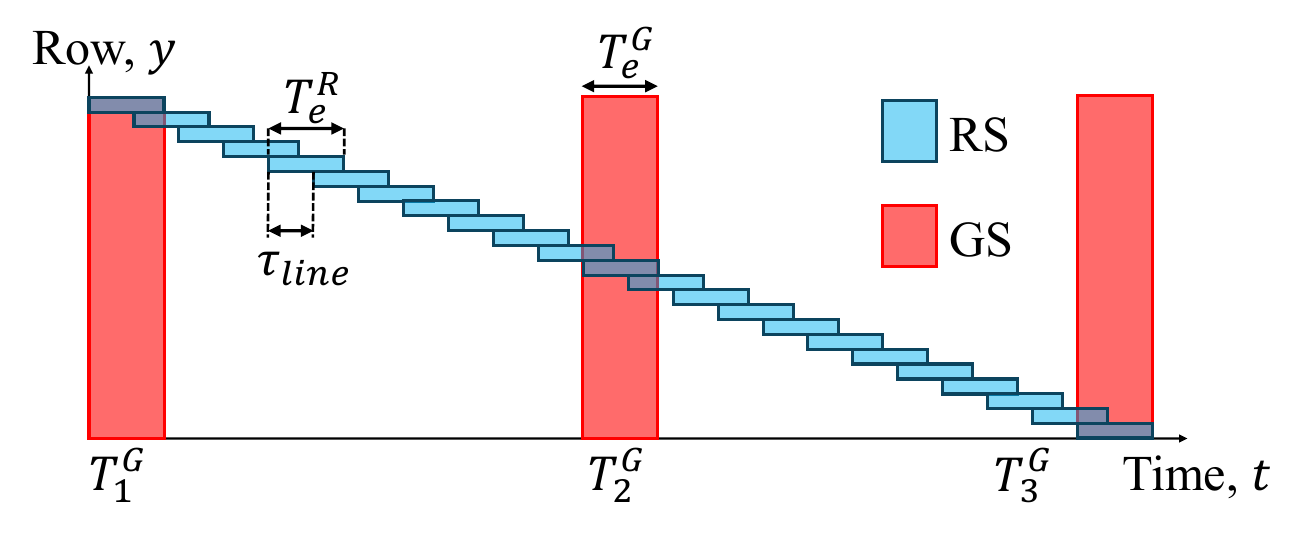}
    \caption{\textbf{Shutter timing diagram.} Blue depicts active RS rows, and red depicts active GS rows. $T^R_e$ is the exposure time for a single RS row, $\tau_\text{line}$ is the lag time between rows, and $T^G_e$ is the exposure time for the GS. At $T_0$, a long RS exposure is triggered, and 3 short GS exposures are triggered at $T^{G}_1$, $T^{G}_2$, and $T^{G}_3$. 
    }
    \label{fig:timing_diagram}
\end{figure}
The timing diagram in Figure \ref{fig:timing_diagram} shows a comparison between global and rolling shutter sensors. The rolling shutter exposes each row of pixels for the same exposure time, $T_{e}^R$ with a short delay, $\tau_{\mbox{line}}$, in the onset of exposure as compared to the previous row. We represent this as the indicator function, \(S_R(x,y,t)  =  \mbox{rect}\left(\frac{1}{T_e^R}\left[t - y\cdot \tau_{\mbox{line}}\right]\right)\).
The delay between rows is typically very short, on the order of microseconds, whereas 
global shutter has a relatively long gap between exposures due to pixel readout.


\subsection{Rolling shutter and Diffusers}
\begin{figure*}[ht!]
    \centering
    \includegraphics[width=1\linewidth]{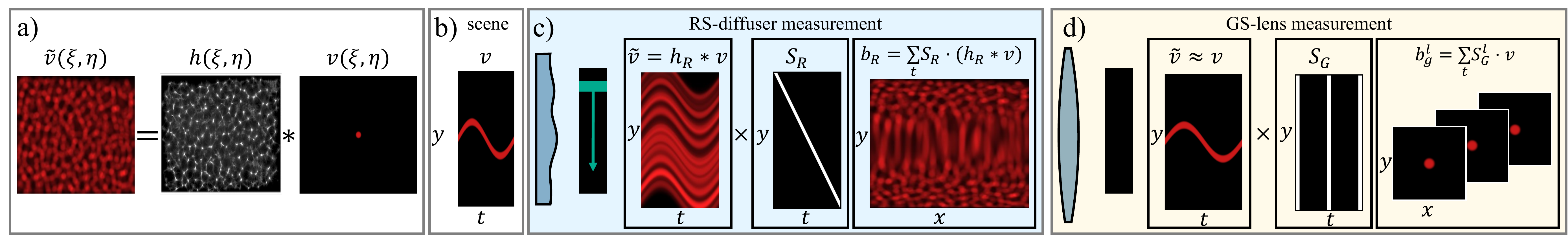}
    \caption{ \textbf{Lensless camera forward model}. (\textit{a}) The intensity arriving at a lensless camera sensor is the 2D convolution of the diffuser PSF $\psf$ and the scene $\obj$.     
     \textbf{Illustrating the capture process of a dynamic scene.} All $y$-$t$ images are slices aligned with the ball's $x$-coordinate.
      (\textit{b}) The scene is a red ball moving sinusoidally in the $y$-direction. The RS-diffuser camera (\textit{c}) measurement $\meas_{R}$, records the dynamic intensity, $\tilde{\obj}$ distributed all over the sensor due to convolution with the large diffuser PSF. This encodes rich spatio-temporal scene information into $\meas_{R}$. The GS-lens camera (\textit{d}) acquires three 2D images, trading temporal information for better spatial detail than the RS-diffuser. This motivates our system design wherein we fuse the two measurement types into a high-speed video. 
     }
    \label{fig:diffuser_RS_GS}
\end{figure*}

The RS arm comprises an RS sensor and a smooth, pseudorandom phase optic called a diffuser. The diffuser maps each scene point to a large, structured PSF, $\psf$.  The intensity arriving at the sensor, $\widetilde{\obj}(x,y,t)$, from an extended scene $\obj(x,y,t)$ is described by convolution~\eqref{eq:discrete_2D_conv}. As illustrated in Figure \ref{fig:diffuser_RS_GS}(a), the large PSF spreads scene information over the entire sensor, which plays a critical role in enabling high-speed video using rolling shutter.

The process of capturing a dynamic scene with a diffuser and RS sensor is illustrated in Figure \ref{fig:diffuser_RS_GS} (c). The single 2D measurement recorded by the RS-diffuser camera, $\meas_R$, is described by substituting $S_R$ and $h$ into~\eqref{eq:discrete_2D_conv} and~\eqref{eq:general_exposure}.  
Because the diffuser distributes the scene intensity values in a structured way over the entire sensor, the $\meas_R$ contains information about nearly all spatial points at each time during RS acquisition. As shown in prior work \cite{antipa2019video, Weinberg2020100000FC}, this enables recovery of sparse video from RS-diffuser measurements. However, these approaches struggle with dense scenes (Fig. \ref{fig:simulation_results} (d) and (e)). We denote the RS-diffuser measurement process for a video with $M \times N$ spatial samples and $K$ frames in matrix-vector form as 
\begin{equation}\label{eq:RS_matrix_form}
    \measVec_R = \Ars \objVec
\end{equation}
where $\Ars: \mathbb{R}^{MNK} \mapsto \mathbb{R}^{MN}$ is the matrix form of \eqref{eq:general_exposure}, $\measVec \in \mathbb{R}^{MN}$ and $\objVec\in\mathbb{R}^{MNK}$ are column-stacked versions of $\meas$ and $\obj$, respectively. The details of $\Ars$ are in \cref{supp:matrix_details}.

\subsection{Global shutter and lens}
The global shutter sensor is coupled with a high quality lens, so we assume its PSF is $h_G(x,y) \approx \delta(x,y)$. However, the lens does have some distortion, and the magnification is not the same as the RS camera, so we perform a coordinate transform on the measurements as described in \cref{supp:coordinate_alignment}.  This allows us to model the intensity arriving at the GS sensor as approximately equal to the scene, $\obj(x,y,t)$. 
In our implementation, we capture three GS exposures at times $T_1^G$, $T_2^G$, and $T_3^G$, spanning the RS capture time. The GS shutter function for the $l-$th GS capture is given by $S_G^l(t) = \mbox{rect}\left(\frac{1}{T_e^G}\left[t - T_l^G\right] \right)$. As illustrated in Figure \ref{fig:diffuser_RS_GS}(d), the combination of GS exposure and a lens produces three frames containing full 2D scene information at the instants the GS sensor was triggered. Note that the GS sensor is blind for most of the video duration due to its slow readout time. The GS measurements are described by
\begin{equation}
    b_G^l(x,y) = \sum_t S_G^l(t) \obj(x,y,t){.}
\end{equation}
We denote in matrix-vector form as
\begin{equation}
    \measVec_G^l =  \Ags^l  \objVec
\end{equation}
where $\Ags^l : \mathbb{R}^{MNK} \mapsto \mathbb{R}^{MN}$ is the matrix describing GS exposure model and $\measVec_G^l$ is the column-stacked version of the $l-$th GS measurement. The collection of all 3 GS measurements, as a vector $\measVec_G \in \mathbb{R}^{3MN}$, is given by
\begin{equation}
    \measVec_G = \Ags \objVec
\end{equation}
where $\Ags = [(\Ags^1)^\intercal|(\Ags^2)^\intercal|(\Ags^3)^\intercal]^\intercal$ is the combined forward model of the three GS captures. Our goal is to compute a high-speed video containing hundreds of frames given only the four frames captured by our two systems: one RS-diffuser capture, and three GS-lens images. With the camera models described above, the high-speed video can be estimated by solving the optimization problem:
\begin{equation}
\label{eq:inverse_problem}
    \widehat{\objVec} = \arg\min_\objVec \|\Afull \objVec -   \measVec\|_2^2
\end{equation}
where $\Afull = \left[\Ags^\intercal | \psi \Ars^\intercal\right]^\intercal$ is a matrix modeling both cameras, and $\measVec = \left[\measVec_G^\intercal |\measVec_R^\intercal \right]^\intercal$ contains all four measured frames in vector form. The parameter $\psi\geq 0$ controls the weight given to the RS measurements. This formulation fuses the GS and RS measurements, allowing us to use GS measurements for estimating 
the spatially high-frequency components of the scene, and RS measurements for temporal frame upsampling. 

This problem is highly underdetermined, so strong video regularization is required to recover the correct video uniquely. While prior work~\cite{antipa2019video} applies 3D total variation (3DTV) on the grid $\mathbf{v}$ as a regularizer, we propose a space-time fusion model (\cref{sec:neural_ST_model}) that allows explicitly regularizing spatio-temporal consistency.




\section{Space-Time Fusion Model Reconstruction}
\label{sec:neural_ST_model}
\begin{figure*}[ht!]
    \centering
    \includegraphics[width=1\linewidth]{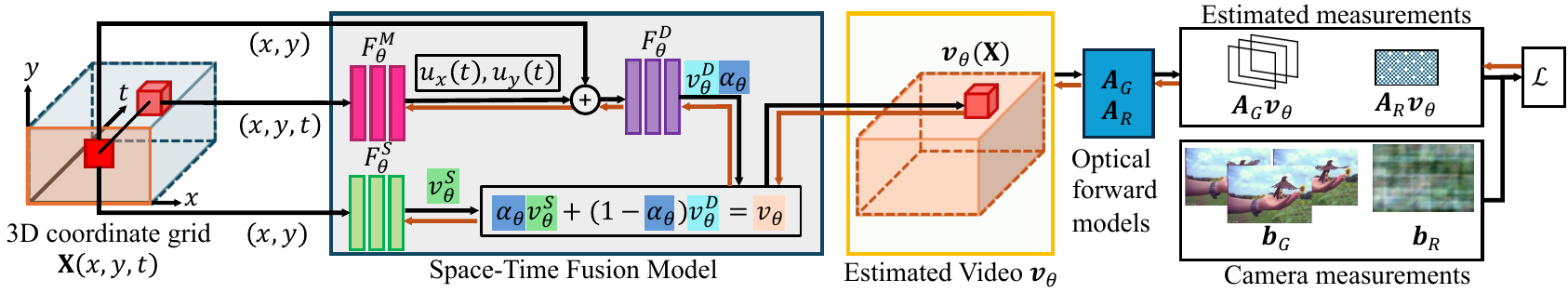}
    \caption{\textbf{Space-Time Fusion Model for compressive video}. Both rolling and global shutter measurements are used to simultaneously update both static and dynamic networks by computing their loss against the estimated measurements from querying the estimated scene $\mathbf v_{\theta}$ and passing it through the optical forward model, $A$.  
    $F^{D}_{\theta}$ takes in a grid of spatiotemporal coordinates while $F^{S}_{\theta}$ only takes in a grid of spatial coordinates. The two outputs are summed together after the alpha map is applied.}
    \label{fig:SIREN_explainer}
\end{figure*}

Our proposed space-time fusion model (STFM) leverages two intrinsic properties of high-speed videos\textemdash static-dynamic decomposition and local spatiotemporal consistency. STFM explicitly factorizes the video into an alpha blend of static (background) and dynamic (foreground) components. Each video component is modeled with a separate INR, see \cref{fig:SIREN_explainer}. We also explicitly model a motion-warping field with a separate INR to locally regularize the spatio-temporal motion in the video. These inductive biases in our design alleviate the ill-posedness of the inverse problem in \eqref{eq:inverse_problem} and significantly improve the reconstruction. We demonstrate this through ablation in \cref{sec:ablations}. We first briefly overview INRs followed by an explanation of our proposed STFM. 



\noindent \textbf{Implicit Neural Representations:  }
INRs are neural networks $F_{\theta} : \mathbb{R}^{P} \to \mathbb{R}^{Q}$, with parameters $\theta$, that provide a continuous approximation to a target function $f : \mathbb{R}^P \to \mathbb{R}^Q$ defined on a $P$-dimensional input domain and producing a $Q$-dimensional signal (e.g., a scalar field or RGB values). When used as coordinate neural networks, they can be trained such that $F_{\theta}(\gamma(\mathbf{x})) \approx f(\mathbf{x}), \forall \mathbf{x}\in \mathbb{R}^{P}$
where $\mathbf{x}$ is the coordinate vector and $\gamma(\cdot)$ denotes the commonly used positional encoding function~\cite{mildenhall2020nerf}, defined in \cref{supp:space-time_fusion}. For brevity we denote $F_{\theta}(\gamma(\mathbf{x}))$ as $F_{\theta}(\mathbf{x})$ in the rest of the text. In our work, we use SIREN~\cite{martel2020neural}, an INR with sinusoidal activations as the representation backbone for our neural space-time model.

\noindent \textbf{Space-Time Fusion Model:  }
We explicitly decompose the video into its static and dynamic components using two separate INRs, $F_{\theta}^S$ and $F_{\theta}^{D}$, respectively, as shown in \cref{fig:SIREN_explainer}. While INRs are continuous functions, our data is measured on a discrete spatiotemporal grid. We therefore evaluate the INRs on a spatiotemporal coordinate tensor $\mathbf{X} \in \mathbb{R}^{MNK\times3} =\left[\mathbf{X}^{xy}, \mathbf{X}^{t}\right]$, where $\mathbf{X}^{xy}\in\mathbb{R}^{MNK\times 2}$ contains the spatial coordinates $(x,y)$ and $\mathbf{X}^{t}\in\mathbb{R}^{MNK \times 1}$ contains the temporal coordinate $t$. We adopt the notation \( F(\mathbf{X}) \) to denote row-wise evaluation: each row of \( \mathbf{X} \) is treated as an input to \( F \). Note that the grid structure arises from the discrete pixel structure of the sensor. More details on $\mathbf{X}$ are in the \cref{supp:space-time_fusion}.

To evaluate the final video color $\mathbf{v}_{\theta}\in \mathbb{R}^{MNK\times 3}$ on the full grid $\mathbf{X}$, we first compute the background RGB color $\mathbf{v}^{S}_{\theta}\in \mathbb{R}^{MNK\times 3}$, which remains constant over time for each spatial location. This is done by evaluating the static INR $F^{S}_{\theta} : (x,y)\mapsto (R,G,B)$ on the spatial grid: $\mathbf{v}^{S}_{\theta} = F^{S}_{\theta}(\mathbf{X}^{xy})$. 
To capture the motion of the dynamic component, we predict a \textit{time-varying} motion warp field  $\mathbf{U}(t)\in \mathbb{R}^{MNK\times2}$ using the motion INR $F^{M}_{\theta}: (x,y,t)\mapsto (u_x(t), u_y(t))$,
 \begin{equation}
\label{eq:motion_network_pred}
    \mathbf{U}(t) = [\mathbf{U}_x(t), \mathbf{U}_y(t)] =F^{M}_{\theta}(\mathbf{X}), 
\end{equation}
where $\mathbf {U}_x(t), \mathbf {U}_y(t) \in \mathbb{R}^{MNK}$ denote the time-varying \(x-\) and \(y-\)direction spatial offsets of the grid $\mathbf X$, respectively.

To compute the dynamic color $\mathbf{v}^{D}_{\theta}(t)$, and transparency $\boldsymbol{\alpha}_{\theta}(t) \in [0,1]$, we first \textit{warp} the spatial input grid $\mathbf X^{xy}$ with the motion field $\mathbf U(t)$. The dynamic INR $F^{D}_{\theta} : (x,y) \mapsto (R,G,B,\alpha)$ is then queried as follows: 
 \begin{equation}
    \left[\mathbf{v}^{D}_{\theta}(t), \boldsymbol{\alpha}_{\theta}(t)\right] = F^{D}_{\theta}(\mathbf{X}^{xy}+ \mathbf{U}(t)).
\end{equation}



The final video is a composite of the static and dynamic colors and is computed as 
\begin{equation}
    \label{eq:final_video}
    \mathbf{v}_{\theta} = \boldsymbol{\alpha}_{\theta}(t)\mathbf{v}^{S}_{\theta} + (1-\boldsymbol{\alpha}_{\theta}(t))\mathbf{v}^{D}_{\theta}(t).
\end{equation}
Alpha compositing the static and dynamic components with a time-varying $\alpha$ helps account for occlusions in STFM. 

\noindent \textbf{Inverse problem:  }
The high-speed video recovery inverse problem in~\eqref{eq:inverse_problem} can now be expressed as
\begin{equation}
    \widehat{\theta} = \arg\min_\theta \|\Afull \objVecTheta -   \measVec\|_2^2
\end{equation}
where we solve for the INR parameters $\theta$ instead of directly optimizing for the spatio-temporal grid. The high-speed video can then be recovered using~\eqref{eq:final_video} with the optimized INR parameters $\widehat{\theta}$. Please see \cref{supp:space-time_fusion} for details.

\begin{figure*}[h]
    \centering
    \includegraphics[width=1\linewidth]{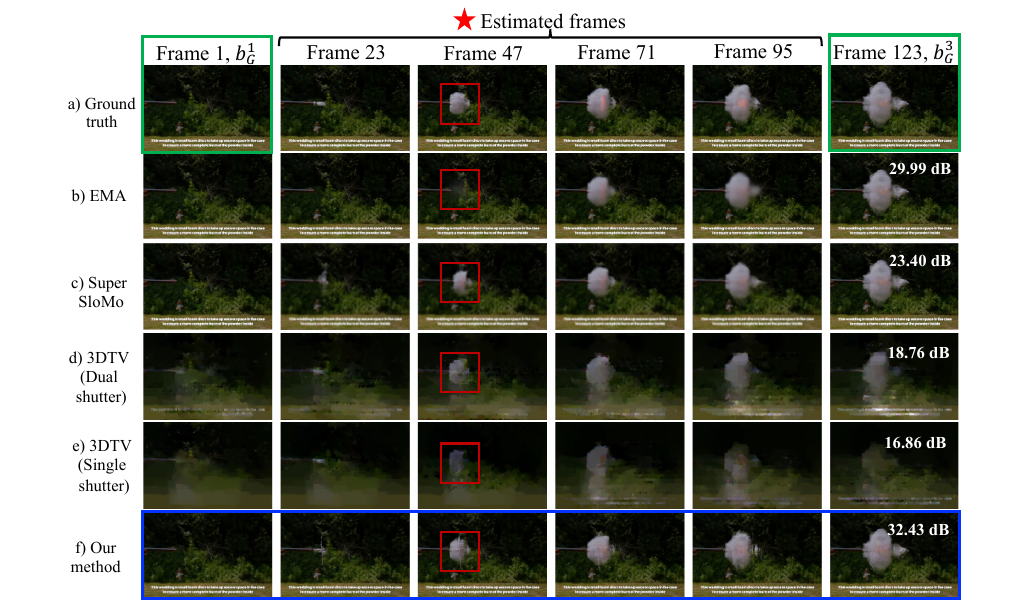}
    \caption{\textbf{Comparing our reconstruction with competing methods on a complex scene.} a) A simulated scene of a smoke plume emerging from a gun barrel (credit: \href{https://www.youtube.com/watch?v=dpr4a89YzE8}{The Slow Mo Guys} ). We compare our method with video interpolators and 3DTV-based methods by calculating PSNR over the entire video. The video interpolators (red inset) b) EMA-VFI \cite{zhang2023extracting} and c) Super SloMo \cite{jiang2018super} fail to recover information present early in the video in the intermediate frames, as they only rely on key-frames (GS) and thus are prone to hallucination. d,e) 3DTV-based methods resolve these details due to the presence of coded RS measurements but have poor reconstruction quality. f) Our method resolves intermediate details with a significantly higher fidelity. We also achieve the highest PSNR calculated over the full video.}
    \label{fig:simulation_results}
\end{figure*}

\noindent \textbf{Regularization:  }
We explicitly model spatial warping through the motion network, regularizing its output $\mathbf{U}$ (we drop $(t)$ for brevity) with anisotropic total variation: 
\begin{equation}
\label{eq:TV_anisotropic}
    \text{TV}_S(\mathbf{U}) = \|\mathbf{U}_x\|_\text{TV} + \|\mathbf{U}_y\|_\text{TV},
\end{equation} where  $\|\cdot\|_\text{TV}$ is the anisotropic 3D total variation semi-norm with temporal weighting factor $\beta$ defined as 
\begin{equation*}
    ||\UVec||_\text{TV} = \sum_{x,y,t} |\nabla_x \UVec| + |\nabla_y \UVec| + \beta |\nabla_t \UVec|, \UVec \in \mathbb{R}^{M\times N \times K}
\end{equation*}
Note that $\mathbf{U}$ is reshaped back to the spatio-temporal grid resolution, before applying anisotropic TV. 
Regularizing the motion fields constrains local dynamics, and significantly improves the reconstruction quality for dense scenes, as demonstrated in \cref{fig:simulation_results}. In comparison, previous works~\cite{antipa2019video,kamilov2017TV} that directly apply 3DTV on the grid $\mathbf{v}$ struggle to regularize dense scenes. 
The resulting optimization objective on incorporating the motion field TV regularization~\eqref{eq:TV_anisotropic}, is given as 
\begin{equation}
\label{eq:invpo}
\begin{aligned}
     \widehat{\theta} &= \arg\min_\theta \left\{ \|\Afull \objVecTheta -   \measVec\|_2^2 + \tau \text{TV}_S(\mathbf{U})\right\}.
    \end{aligned}
\end{equation}
For experimental data, we slightly modify \eqref{eq:invpo} to also optimize for the relative white balance between the GS and RS measurements, see \cref{supp:coordinate_alignment} for details. 
We defer the implementation details, including the hyperparameters $\tau,\beta, \psi$, and INR architecture in the \cref{supp:prototype detail}.  



\begin{figure}[ht]
    \centering
    \includegraphics[width=8cm]{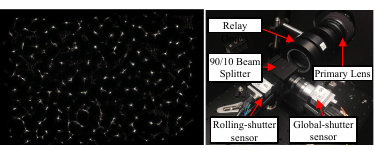} 
    \caption{\textbf{RnGCam Experimental Setup.} 
    \textit{Right:} RnGCam hardware consists of an RS arm (RS sensor and optical diffuser)and the GS arm (GS sensor and lens). We use a beam splitter supplemented with relay optics to ensure optical consistency between the two arms. All these components are placed inside a light-tight box.
    \textit{Left:} PSF captured for the RS arm.}
    \label{fig:experimental_setup}
    \vspace{-12pt}
\end{figure}
\begin{figure}
    \centering
    \includegraphics[width=1\linewidth]{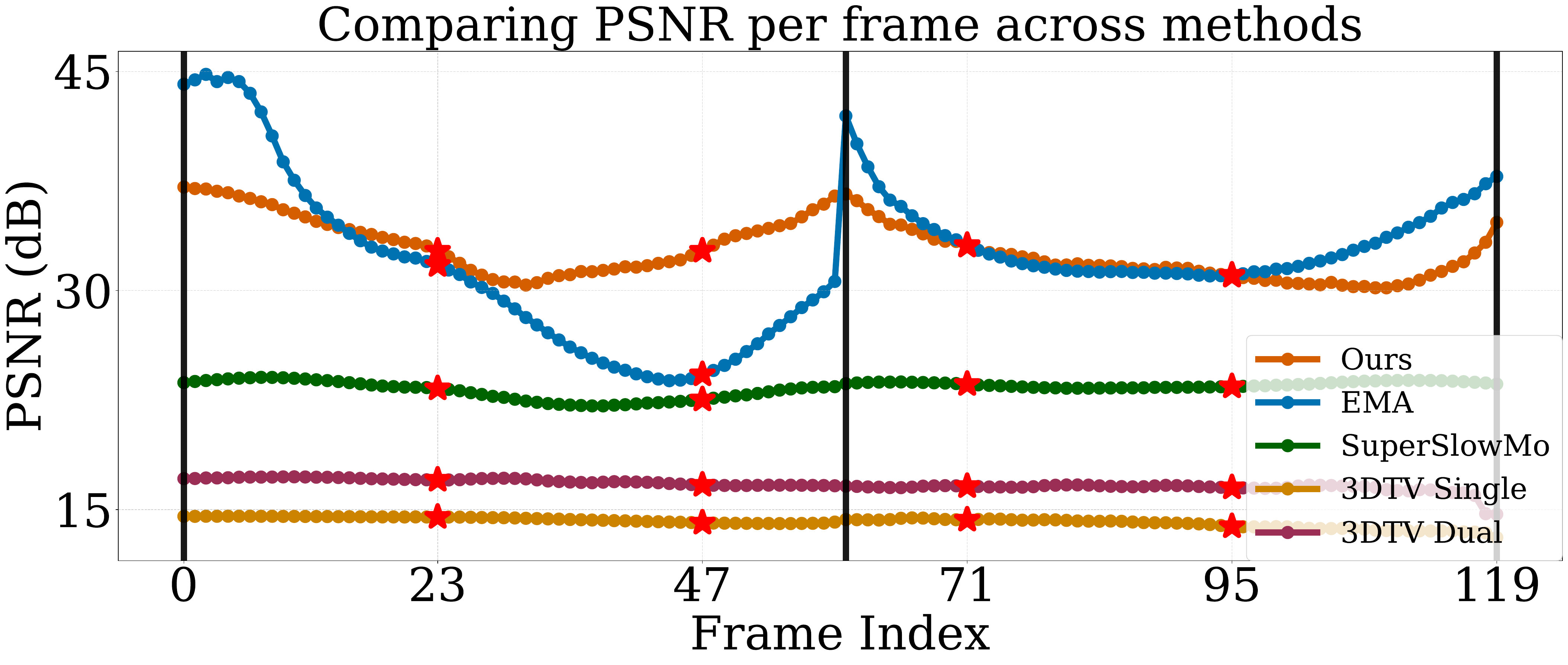}
    \caption{\textbf{Comparing per-frame PSNR for all methods}. Video interpolators achieve high PSNR near the three input GS frames (start, mid, end) indicated by the black vertical lines, but show degradation for intermediate frames. Our method is more temporally stable, achieving the highest PSNRs on intermediate frames (excluding GS frames). We show results on the scene in \cref{fig:simulation_results}, with the stars corresponding to the estimated frames in \cref{fig:simulation_results}.  
    \vspace{-18pt}
    }
    \label{fig:psnr-perframe}
\end{figure}

%% file: sec/6_results.tex
\section{Results}
\label{sec:results}

In this section, we compare the performance of our method relative to video interpolators and 3DTV \cite{antipa2019video} in simulation, and demonstrate our reconstructions at 4,800 fps on real experimental data obtained from our hardware prototype.


\subsection{Simulation results}
\begin{figure}[ht]
    \centering
    \includegraphics[width=\linewidth]{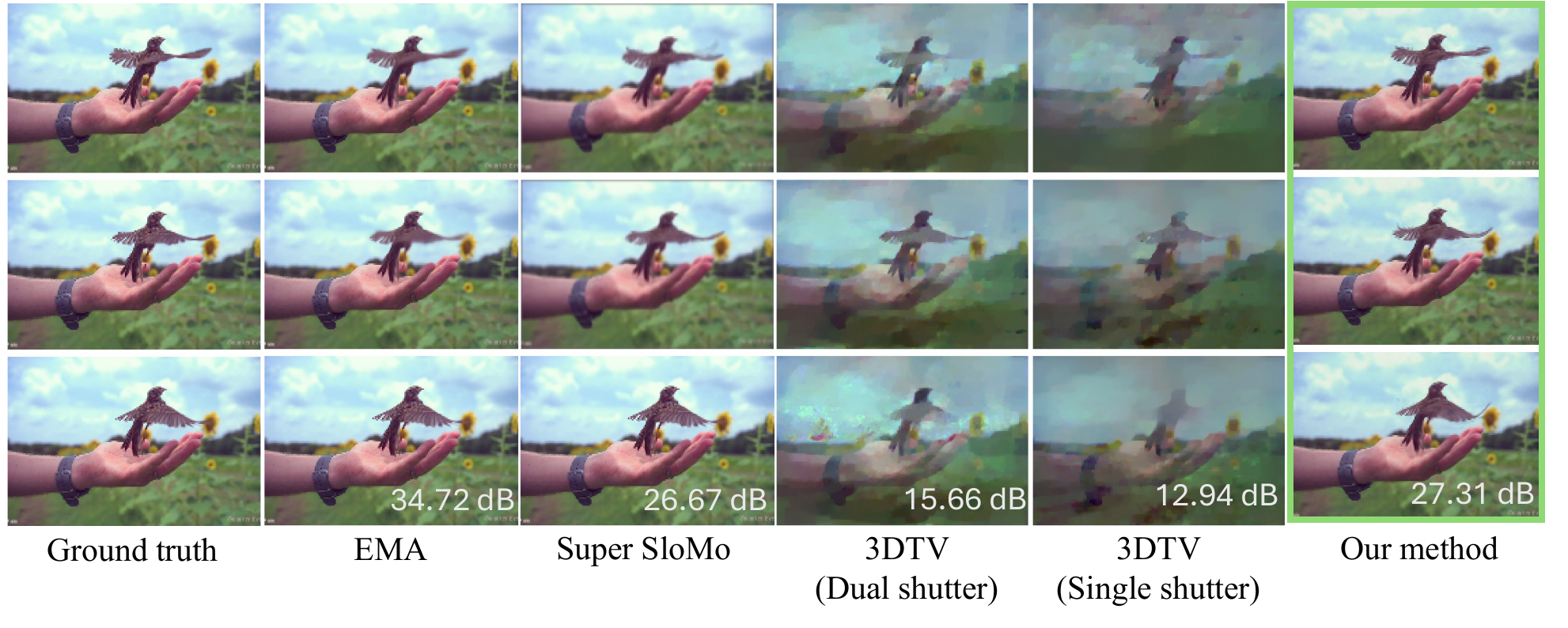} 
    \caption{\textbf{Failure Case.} We recover high-frequency details better than 3DTV-based methods, but video interpolators perform comparably (SuperSloMo) or better (EMA). This is expected since video interpolators are well suited for scenes with sparse and simple motions, such as the shown bird wing flap (credit: \href{https://www.youtube.com/watch?v=iVKdU_OmsaU}{SmarterEveryDay}).} \vspace{-12pt}
    \label{fig:bird}
\end{figure}

We compare our reconstructed results with several data-driven video interpolators \cite{zhang2023extracting, jiang2018super, jin2023unified} and reconstruction with 3D total variation (3DTV) regularization adopted by previous papers \cite{antipa2019video, Weinberg2020100000FC} in \cref{fig:simulation_results}, for a scene with complex motion patterns. 
We use the 3 GS frames as input for the video interpolators and solve for 60 intermediate frames. For the single and dual shutter 3DTV reconstructions, the inputs are the spatially multiplexed rolling shutter measurements. Classical TV-based methods perform poorly for scenes with dense backgrounds due to the lack of strong motion and smoothness priors see~\cref{fig:simulation_results}.
Video interpolators tend to perform better or comparable to our method on scenes that resemble their training data, such as simple and sparse motion like in~\cref{fig:bird} with a natural image background. For more complex motions (\cref{fig:simulation_results}), we outperform all the baselines. The combination of high-frame-rate RS measurements with STFM regularization enables our method to accurately recover intermediate frame details in scenes with complex motion. 
In \cref{sec:ablations}, we include ablations to examine the contributions of different components of both the STFM model and our measurements.

\begin{figure*}
    \includegraphics[width=1\linewidth]{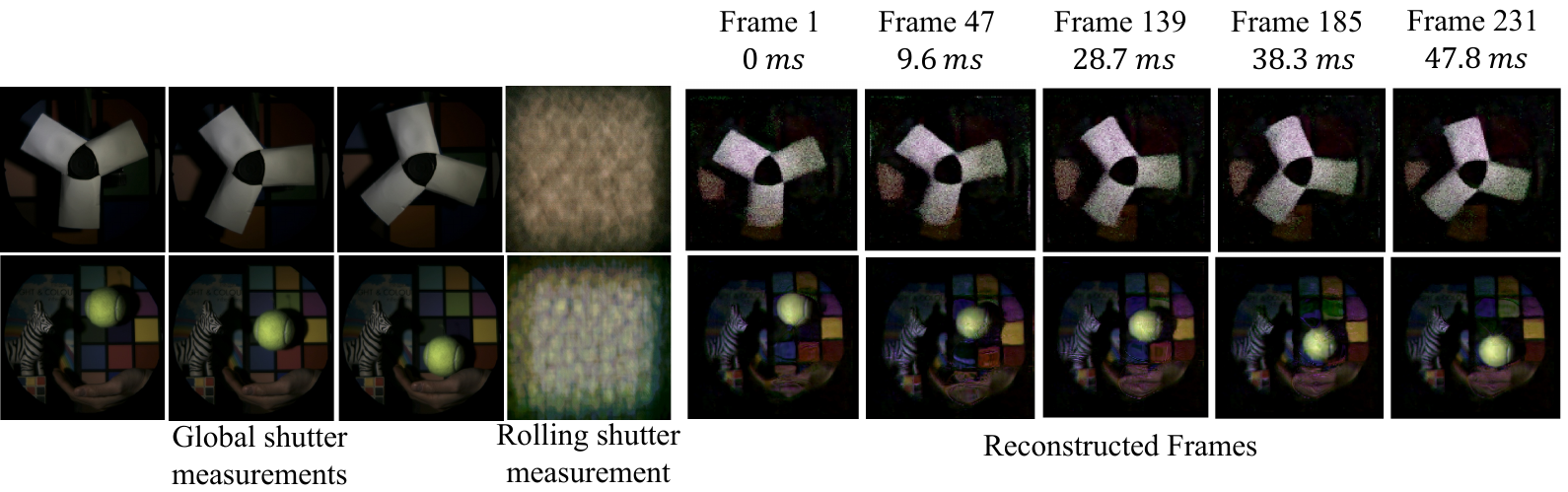}
    \caption{\textbf{Experimental results from our hardware setup.} We show the 3 GS  (\textit{left}) and RS measurements (\textit{middle}) for each scene. We show a subset of the 231 reconstructed frames (\textit{right}) at $4807$ Hz. We recover the static and dynamic components with high fidelity.}
    \label{fig:exp-real}
\end{figure*}

\subsection{RnGCam Hardware Setup}
\label{subsec:hardware}

We designed and built the RnGCam prototype, (\cref{fig:experimental_setup}), illustrated in \cref{fig:Pipeline}. 
Our setup utilized a relay optical system and a 90/10 beam splitter followed by a primary lens to simultaneously collect the same scene using RS with diffuser and GS with lens.
We use set exposure times $T^G_e, T^R_e $ of GS and RS sensors to $650 \mu s$. Our RS sensor has a line time of $13 \mu s$. 
Due to memory limitations for the coordinate network grid, we downsample our working grid by $16$ in space and time, resulting in an effective line time of $208 \mu s$ (4,807 fps).
Details about the prototype implementation are in ~\cref{supp:prototype detail}. In contrast to previous RS-diffuser works ~\cite{antipa2019video}, our setup does not require a dual shutter sCMOS camera, which costs $>\$20k$ USD, and produces better results.
\subsection{Experimental results}
In Figure \ref{fig:exp-real}, we demonstrate the ability of our prototype to resolve high-speed dynamics with a dense background component. We record a scene of a spinning propeller with a checkered background (top), and a scene of a tennis ball being caught by a hand, with complex background (bottom) containing high-frequency details, e.g., the zebra. 

\begin{figure}
    \centering
    \includegraphics[width=0.95\linewidth]{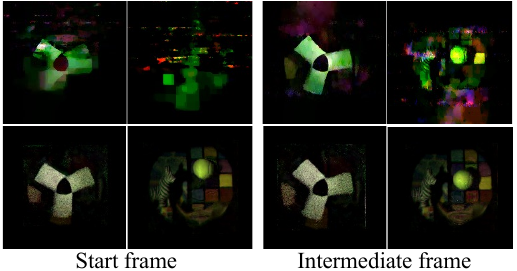}
    \caption{\textbf{3DTV failure with single shutter}. 3DTV for a single shutter (\textit{top}) fails at early  frames due to blind spots. Reconstructions at intermediate frames still contain artifacts. In contrast, our method (\textit{bottom}) recovers these frames at high fidelity.}
    \label{fig:3dtv_exp}
\end{figure}

Our method temporally resolves the spinning propeller (\cref{fig:exp-real}, top row), while simultaneously estimating and blending the occluded color checker in the background with the propeller blades. We also resolve the motion of the tennis ball together with the complex background scene (\cref{fig:exp-real}, bottom row). In \cref{fig:3dtv_exp}, we show that 3DTV performs poorly on the start frame due to the time-varying field of view \cite{antipa2019video}; intermediate frames also contain strong artifacts and poor detail.

%% file: sec/7_limitations.tex
\section{Limitations}
\label{sec:limitations}
We assume the capture setup to be static, and that the scene can be factorized into a static background and dynamic foreground. 
Our proposed STFM relies on a warping assumption to model inter-frame motion. As a result, it struggles with scenes where objects appear suddenly, violating the spatio-temporal consistency assumption. We use SIREN in SFTM, using more recent video-specific INRs~\cite{Saragadam_2023_CVPR} might further improve video reconstruction quality. SFTM fits to all data, and GS frames are not hard constraints.

Our method requires scenes to be well-lit due to the RS-Diffuser sensor's low light throughput, caused by the diffuser's large sensor footprint. We are limited by GPU memory constraints and the small pixel size of commercially available rolling shutter sensors, restricting us to lower spatial and temporal resolutions. In the absence of these constraints, our method can theoretically achieve $77,000$ fps. 





%% file: sec/8_future_work.tex
\section{Conclusion}
\label{sec:conclusion}
We demonstrated compressive high-speed video recovery at an effective frame rate of $4.8$ kHz with a dual camera setup consisting of low-cost consumer-grade sensors.  
Our prototype \textit{RnGCam} is 10x cheaper than existing compressive video recovery setups~\cite{antipa2019video}. 
RnGCam consists of an RS-diffuser arm that captures high temporal detail and a GS arm that captures high spatial detail. 
Fusing these complementary measurements lets us recover high-speed videos with dense backgrounds. 
Previous methods rely on much more expensive hardware but still fail in this scenario. 
For fusing GS-RS measurements, we proposed an INR-based space-time fusion model, which explicitly imposes static-dynamic factorization on the video and also models spatio-temporal warping.
These inductive biases significantly improve the quality of our recovered high-speed videos. 
We evaluate our method on simulated and real data captured with RnGCam. 
We outperform an existing computational imaging method ~\cite{antipa2019video} and also demonstrate superior performance over data-driven video interpolators for scenes with complex motion. 

\newpage
\noindent\textbf{Acknowledgements} This work was supported by a Kavli Institute for Brain and Mind Innovative Research Grant. 


%% file: sec/X_suppl.tex
\newpage
\appendix
\setcounter{equation}{0} 
\setcounter{figure}{0}
\setcounter{page}{1}
\maketitlesupplementary

This appendix material is organized as follows. In \cref{supp:coordinate_alignment}, we implement several processing steps to properly apply the proposed method to the hardware data, including image size alignment, white balance correction, and memory limitations during reconstruction.
In \cref{supp:prototype detail}, we provide details on the prototype for creating an RnG Cam. 
We present an affordable method for making a random diffuser, along with specifications on the optical system. 
This includes information on achieving shift invariance in the optical system and calibrating the point spread function (PSF). 
Additionally, we will discuss the proper setup for the system time settings and the neural space-time model.
We also present additional results, including ablations, and full model results based on experimental data in \cref{supp:misc}.

\section{Handling different sensor sensitivities and resolution}
\label{supp:coordinate_alignment}
\subsection{Aligning rolling and global shutter images}

To align measurements from the two sensors, we simultaneously capture a static calibration image on both camera. We deconvolve the diffuser-coded image on the rolling shutter to obtain the scene from the RS viewpoint. The calibration measurement from the GS sensor is then aligned with the deconvolved RS scene by aligning two features in the scene with a scale and rotate transformation. 

\subsection{White balance correction between global and rolling shutter measurements}
The GS and RS sensors have different sensitivities per channel. Additionally, they have different optical elements in front of the sensors. The GS sensor has a lens, while the RS has a random optical diffuser. 

To calibrate the two sensor white balance and energy levels, we predict the per-channel color correction coefficients using the static INR as 3 extra outputs $\lambda_r, \lambda_g, \lambda_b$ which we represent as a matrix  
\begin{equation}
    \mathbf\Lambda = \begin{bmatrix}
    \lambda_r&0&0\\0 & \lambda_g & 0 \\0 & 0& \lambda_b
\end{bmatrix}.
\end{equation}
We slightly modify the optimization objective in eq. 14 in the main text to incorporate the correction factor $\mathbf\Lambda$ as follows: 

\begin{equation}
\label{eq:final_loss_lambda}
    \widehat{\theta} = \arg\min_\theta \|\Ags \objVecTheta -   \measVec_G\Lambda\|_2^2 + \psi \|\Ars \objVecTheta - \measVec_R\|_2^2 + \tau TV_S(\mathbf{U}).
\end{equation}


\subsection{Memory limit, evaluating subset of INR}

In our current implementation, there is a field stop, which makes the image 0 outside the region defined by the field stop. For memory efficiency, and reducing the extent of downsampling, we evaluate the model only inside the field stop region, containing nonzero intensities. 

\subsection{Modeling details}\label{supp:matrix_details}
Equation \eqref{eq:general_exposure} describes the forward model of a general camera with static psf $\psf(x,y)$ and shutter function $S(x,y,t)$. Substituting the discrete implementation of linear convolution, \eqref{eq:discrete_2D_conv}, into \eqref{eq:general_exposure} yields
\begin{equation*}
    \meas(x,y) = \sum_T S(x,y,t) \crop \left[ \pad \left( \psf(x,y) \right) \stackrel{(x,y)}{\circledast} \pad \left( \obj(x,y,t)\right ) \right]
\end{equation*}
We represent this compactly as a matrix-vector multiply in \eqref{eq:RS_matrix_form}. The system matrix can be conceptualized as 
\begin{equation*}
\mathbf{A = \Sigma}\mbox{diag}(\mathbf{S})\mathbf{C}\mathbf{F}^{-1}\mbox{diag}(\mathbf{FPh})\mathbf{F}{.}
\end{equation*}
Here, $\mathbf{\Sigma}$ is the matrix for summation over time. Pointwise multiplication by the shutter function, $S$, is described by $\mbox{diag}(\mathbf{S})$, which is a diagonal matrix comprised of the column-stacked shutter indicator, denoted $\mathbf{S}$. $\mathbf{F}$ is the 2D Discrete Fourier Transform matrix, $\mathbf{\psf}$ is the column-stacked point spread function (PSF), $\mathbf{P}$ and $\mathbf{C}$ are the matrix forms of zero-padding and cropping, respectively. Note that, in practice, we implement the camera model using operators; matrices are used only for compact notation here. 



\section{RnG Cam Prototype Detail}
\label{supp:prototype detail}
\begin{figure}
    \centering
    \includegraphics[width=1\linewidth]{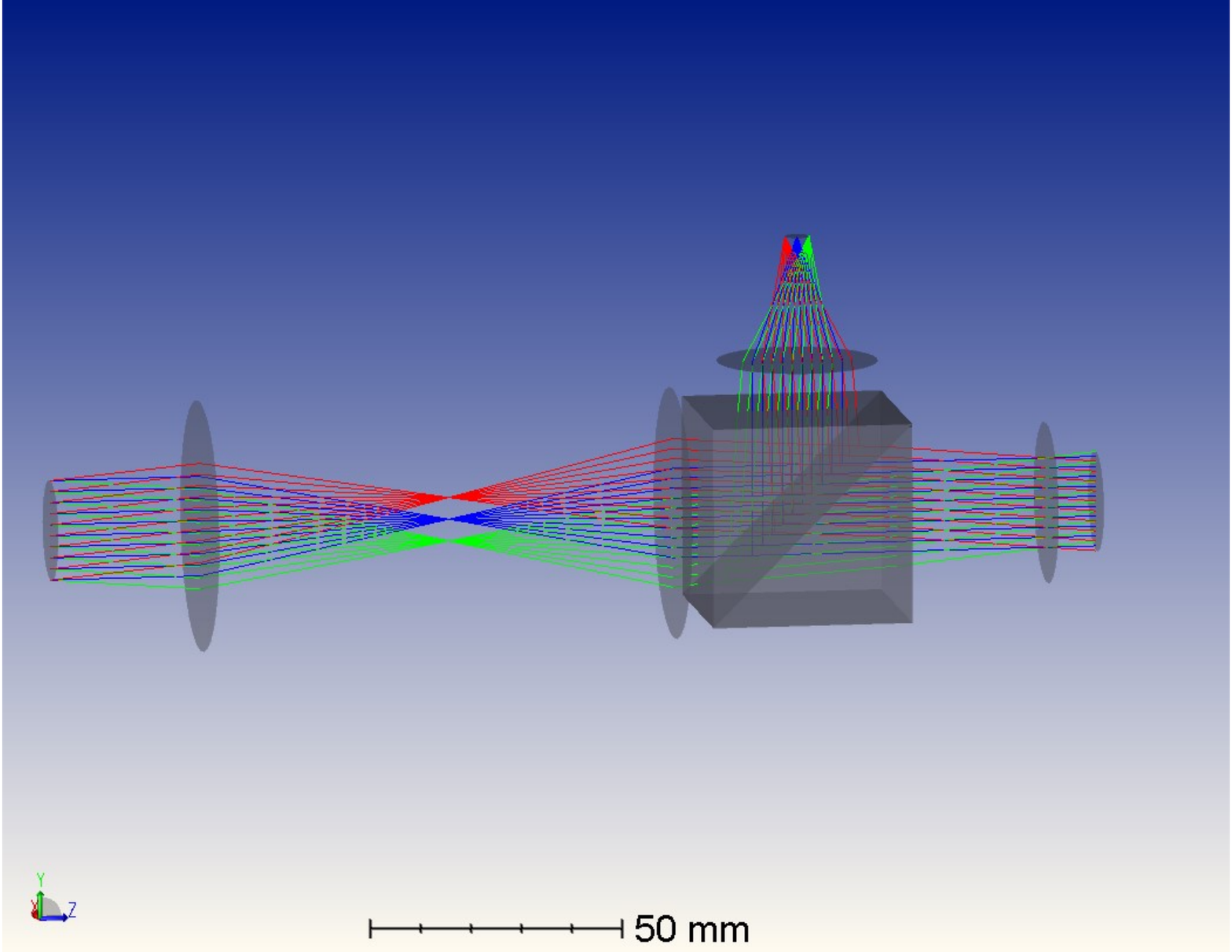}
    \caption{3D Zemax design}
    \label{fig:3dzemax_design}
\end{figure}
The overview of the RnG Cam is illustrated in Fig.~\ref{fig:3dzemax_design}. In the following subsections, we will discuss the importance of a well-designed diffuser and relay lens in achieving system shift invariance and extending the bandwidth limit of the measurement.
\subsection{Diffuser Design and Manufacturing}
A diffuser must fulfill three essential requirements: First, its PSF should create a random pattern to avoid periodicity. Second, it should cover as much of the sensor area as possible to enhance the bandwidth of a snapshot. Third, the feature size needs to be small enough to be sensitive to motion.
To achieve this, we create randomly positioned unifocal lenslets using 9/16-inch ball bearings, resulting in a focal length of approximately 28 mm. With this focal length, we can place the diffuser against the camera housing to generate a sharp point on the sensor when the incident light is collimated.

The diffuser we used for the experiment is both low-cost and easy to make. 
First, randomly dent the polished aluminum block using a 9/16-inch stainless steel ball. Then, apply optical epoxy to cover the dented area. 
Next, place a clear cover slide over the epoxy and cure it using the appropriate wavelength of UV light. 
Finally, the diffuser is created by peeling off the cover slide from the aluminum block.

\subsection{Shift Invariant Imaging System}
\begin{figure}
    \centering
    \includegraphics[width=1\linewidth]{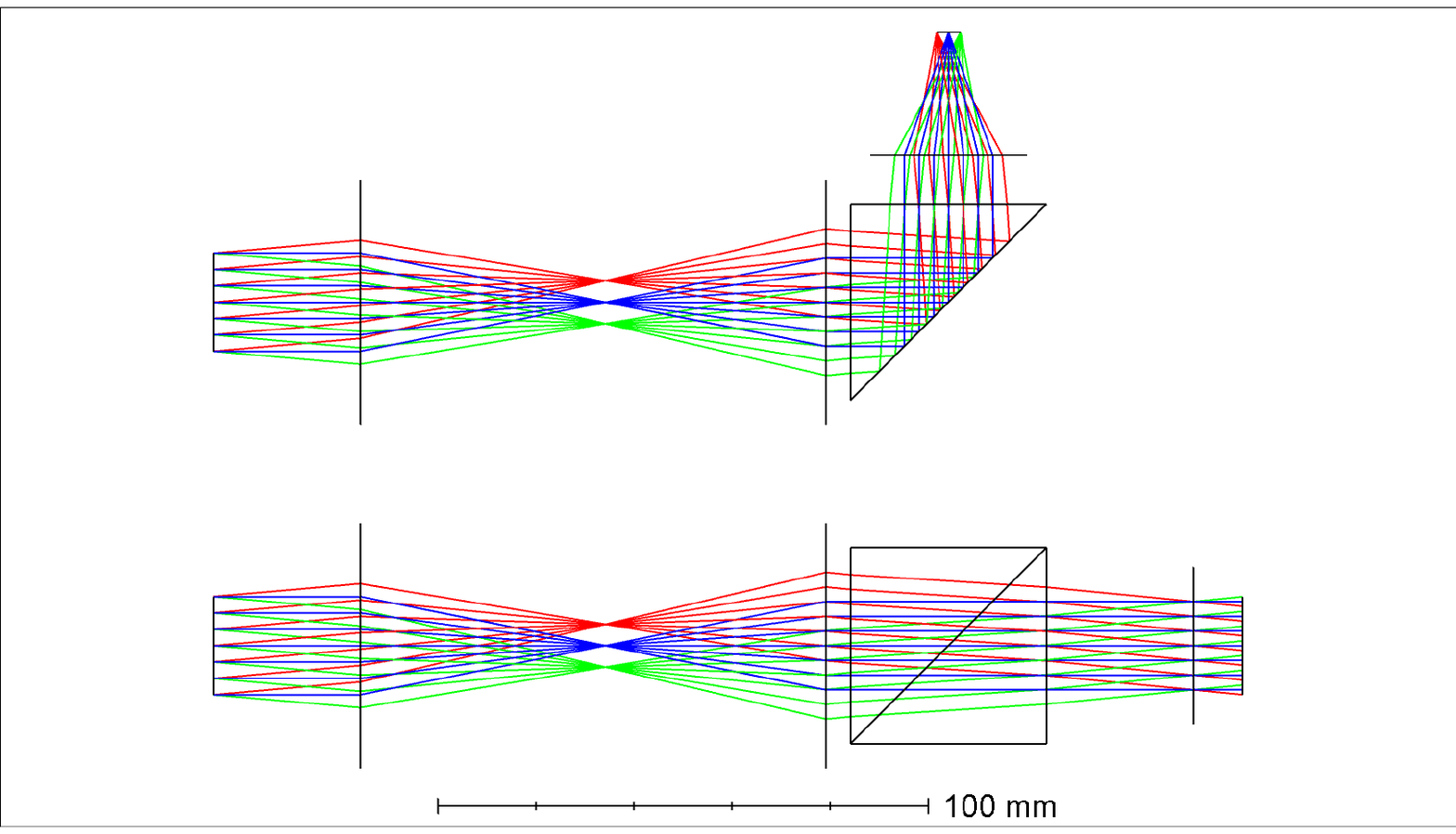}
    \caption{2D Zemax diagram}
    \label{fig:2dzemax_design}
\end{figure}

In the inverse problem of the diffuser cam, it is essential to position the diffuser at the aperture stop of the optical system to ensure that the problem remains shift invariant to simplify the deconvolution and system calibration.
Our setup includes two sensors that utilize the same camera lens as the primary lens, but the aperture stop is inaccessible because it is located within the camera lens.
Thus, the follow-up collimated lens serves two purposes: one is to collimate the light for the beam splitter to reduce aberrations, and the other is to reimage the stop plane to a physically accessible location as the bottom diagram in Fig.\ref {fig:2dzemax_design} shows.

\subsection{Optical System Details}
We present the prototype of the system in Fig.~\ref{fig:experimental_setup}. 
The system uses global- and rolling-shutter cameras to capture the same scene simultaneously through a primary lens followed by a relay optical system with a beam splitter at the conjugate plane. 
The primary lens is a Sigma 50mm EX DG HSM Lens with f-number 1.4, and a field stop is set at the focal plane to control the field of view.
Then, two 2-inch 98 mm focal length doublets with effective focal lengths around 45 mm collimate the light and image the aperture of the primary lens to a physically accessible plane where the diffuser will be located to avoid vignetting.
The distance between the two doublets' last surface and the pupil's image is designed to be sufficient to fit in a 1 inch visible light beam splitter, which has 90/10 uneven energy distribution. 
The RS arm requires more energy because of spatial multiplexing. 
Therefore, the higher intensity arm has a 1" format RS camera (Basler ace acA5472-17uc, IMX183) with a diffuser containing random microlenses with a 9/16-inch radius and an effective focal length of around 28mm. 
The weaker intensity arm of the beam splitter has one 1/1.2" format global shutter camera (Basler ace acA1920-155uc, IMX174) with a Fujinon 25mm 1.4 f-number machine vision lens to form the static reference image. 

A function generator syncs cameras with a hardware trigger to start simultaneously, triggering the subsequent three global shutter frames between a RS frame.
\subsection{PSF calibration}

The point spread function of the system, $h$, shown in Figure \ref{fig:experimental_setup} (a) is experimentally obtained by shining a point light source to the main lens of the system shown in Figure \ref{fig:experimental_setup} (b). 
\subsection{System Timing}
The downsampling factor of 16 in space and time was chosen because of a combination of memory limitations and the very high resolution of the RS sensor ($T,W,H \approx 3648 \times 5472 \times 3648$). Because our effective frame rate is limited by the downsampling we have to fit the coordinate network to our machine (48GB NVIDIA A40), a lower resolution RS sensor with a similar line time could get us to around $77,000$ fps.

\subsection{Space-time fusion model implementation details}\label{supp:space-time_fusion}
In our implementation, we instantiate the time-varying SIREN $F_{\theta}^{D}$ with 3 hidden layers, and 128 hidden features, the static SIREN $F_{\theta}^S$ and motion SIREN $F_{\theta}^M$ with 2 hidden layers and 32 hidden features. We apply a non-negativity constraint on the outputs of $F_{\theta}^S$ and $F^{D}_{\theta}$ and apply a sigmoid activation to ensure that $\alpha \in [0,1]$.

To increase the ability of the scene representation to represent higher frequency features \cite{DBLP:journals/corr/abs-2006-10739}, we apply positional encoding $\gamma(\cdot)$ to the 3D coordinate inputs $(x,y,t) \in \mathbf X$, where 
\begin{equation}
\gamma(x) = (x, \cos{(2^i\pi x)}, \sin{(2^i \pi x)}, \dots) \text{, for } \textit{i} = 0, \dots, L-1.
\end{equation}
$L \in \mathbb{Z}^{+}$ is a tunable parameter. Larger values of $L$ increase the ability of the network to represent high-frequency information. However, as seen in \cite{9924573, cao2022dynamic} large $L$ values may introduce high-frequency distortions and overfitting in the final reconstructions. For all the subnetworks, $F_{\theta}^{M}, F_{\theta}^{D}, F_{\theta}^{S}$, we consider $L$ as a tunable parameter. 

$F^M_\theta$ and $\beta$ is an additional weight the temporal total variation sparsity. (We found that $\beta$ values from 10 - 10000 yielded the best results, depending on the scene). 


The estimated measurements are used to compute the mean squared error with $b_G$ and $b_R$ and we minimize this loss with respect to the parameters of the coordinate networks. We minimize this using the Adam optimizer with learning rate 0.5e-5. We ran all experiments for a fixed number of iterations, with average total runtimes of approximately three hours.

\section{Miscellaneous Results}
\label{supp:misc}

\subsection{Ablations on global shutter and motion regularization}
\label{sec:ablations}

\begin{figure*}[ht]
    \centering
    \includegraphics[width=15cm]{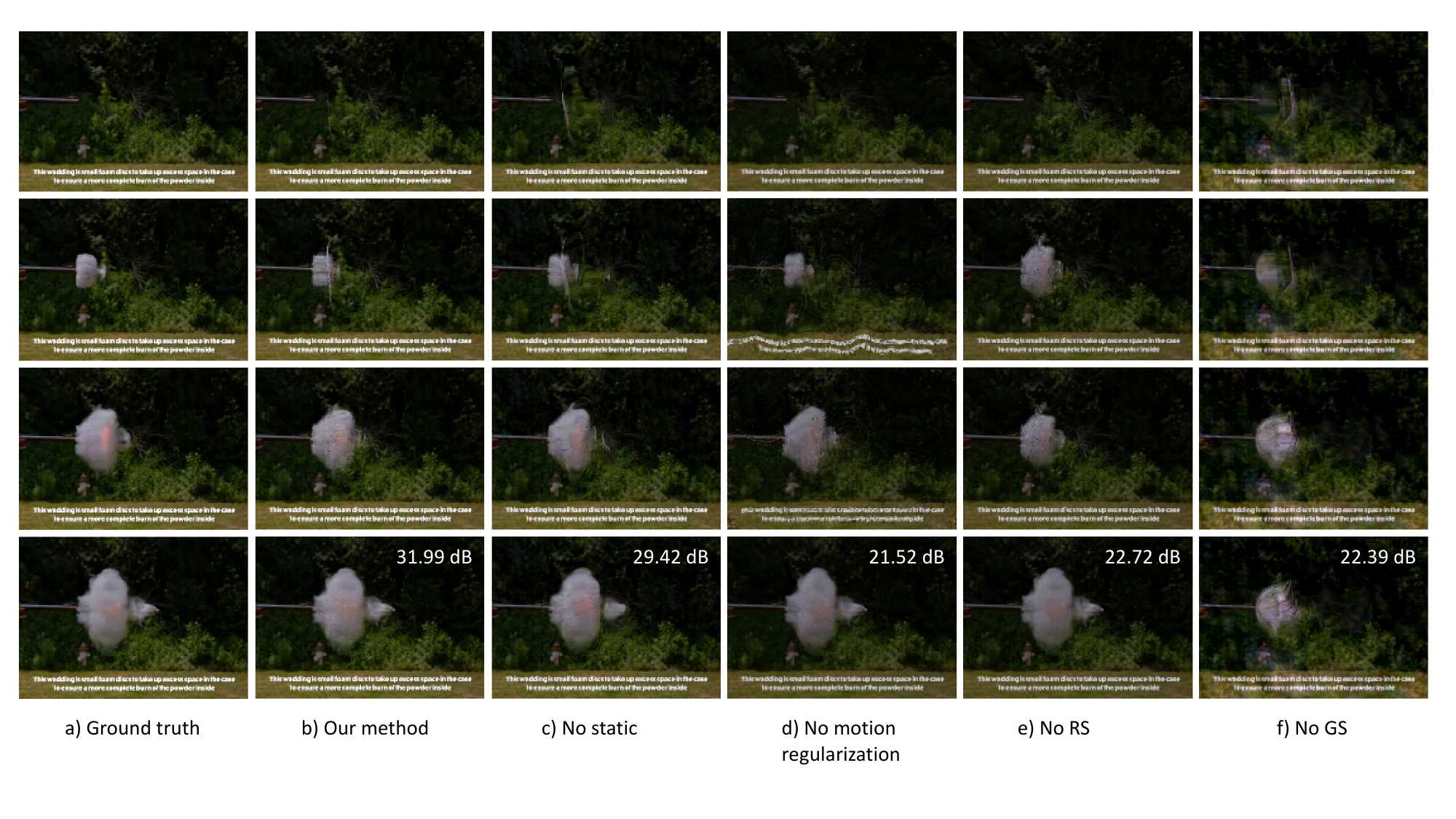} 
    \caption{\textbf{Ablations on bullet scene.} We compare our method (b) to ablations removing different components of our design: c) we remove the static INR; d) without motion regularization; e) without RS measurements; f) without the three GS captures.}
    \label{fig:ablations_visualization}
\end{figure*}

In this section, we perform ablations to examine the contributions of different components of both the Neural Spacetime model and our design including both RS and GS  measurements. The results are summarized in \cref{tab:ablation table} and \cref{fig:ablations_visualization}.

Without motion regularization we observe, for example, distortion in the appearance of the captions at the bottom of the frames.
Without RS measurement, we do not correctly recover the initial appearance of smoke emerging from the barrel.
Without the GS measurement, we do not correctly recover the full puff of smoke at the final frames. We demonstrate that all the components in our design together contribute to recovery complex motion with a dense background.

\begin{table}
    \centering
    \setlength{\tabcolsep}{2pt} 
    \begin{tabular}{|c|c|c|c|c|c|} \hline 
         label&  GS&  RS&  Motion Reg&  Static network& PSNR\\ \hline 
         Our method&  $\checkmark$&  $\checkmark$&  $\checkmark$&  $\checkmark$& \textbf{31.99 dB}\\ \hline 
         no static&  $\checkmark$&  $\checkmark$&  $\checkmark$&  $\times$& 29.42 dB\\ \hline 
         no motionreg&  $\checkmark$&  $\checkmark$&  $\times$&  $\checkmark$& 21.52 dB\\ \hline 
         no RS&  $\checkmark$&  $\times$&  $\checkmark$&  $\checkmark$& 22.72 dB\\ \hline 
         no GS&  $\times$&  $\checkmark$&  $\checkmark$&  $\checkmark$& 22.39\\ \hline
    \end{tabular}
    \caption{Ablation test on bullet scene. See \cref{fig:simulation_results}. We test the effects of removing individual parts of our model, including the static network, regularization on the motion field, rolling shutter measurements, and global shutter measurements. We show that a combination of every component yields the best reconstruction. }
    \label{tab:ablation table}
\end{table}

\begin{figure*}[ht]
    \centering
    \includegraphics{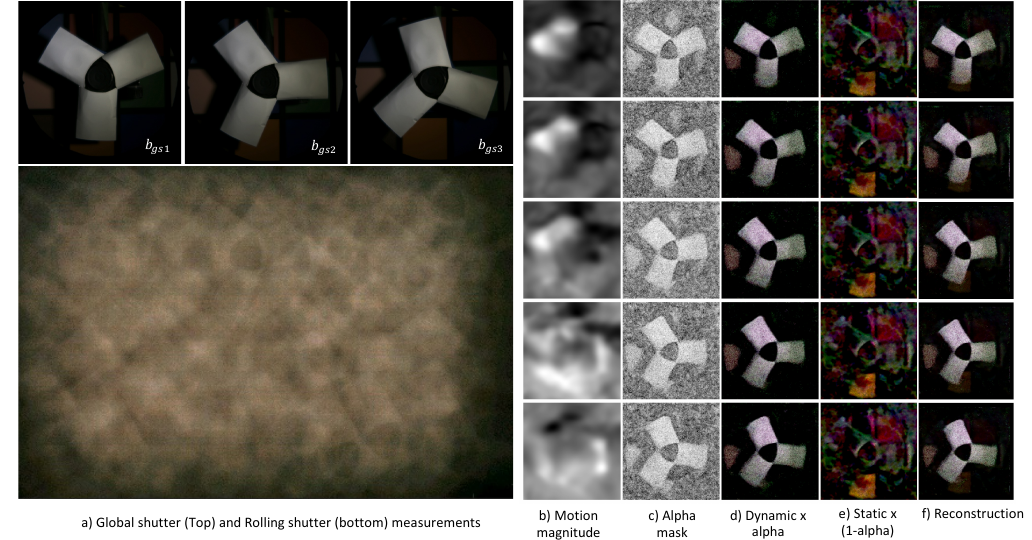}
    \caption{\textbf{Visualizing intermediate components of our method on an experimental scene.} \textbf{1) Reconstructions of a spinning propeller} \textbf{(a)} Spatiotemporally encoded RS measurements (bottom), and the 3 global shutter measurements acquired over the same period (top). \textbf{(b)} Magnitude of the motion encoding from the motion network. \textbf{(c)} Time-varying alpha mask used to blend estimated static (e) and dynamic (d) scenes. \textbf{(d)} Dynamic estimate multiplied by alpha mask. \textbf{(e)} Static scene (contrast stretched for visualization) multiplied by (1-alpha mask). \textbf{(f)} Full scene reconstruction. \textbf{2) Reconstructions of a tennis ball leaving hand.} We demonstrate that our system is able to simulatenously resolve both the dense background, and the dynamics of the tennis ball.}
    \label{fig:experimental_results_propeller supplement}
\end{figure*}

\subsection{Full model results on experimental data}
We visualize the full reconstruction of our network, with all of its intermediate components in \cref{fig:experimental_results_propeller supplement}.
We present the magnitude of the motion encoding in b), the time-varying alpha mask which enables blending of the dynamic motion d) with the static background e) resulting in the reconstructed scene f).